\documentclass[12pt,usenatbib]{mn2e}
\usepackage{graphicx}
\usepackage{epsf}
\usepackage{lscape}
\usepackage{longtable}
\usepackage{rotating}

\newcommand{\lesssim}{\la}
\newcommand{\aap}{A\&A}
\newcommand{\araa}{ARA\&A}
\newcommand{\apj}{ApJ}
\newcommand{\apjl}{ApJ}
\newcommand{\apjs}{ApJS}
\newcommand{\mnras}{MNRAS}
\newcommand{\aj}{AJ}
\newcommand{\apss}{Ap\&SS}
\newcommand{\pasp}{PASP}

\newcommand{\nat}{Nat}

\begin{document}

\title[The Evolution of Stellar Mass]{The Evolution of Stellar Mass and the Implied Star Formation History}

\author[Stephen M. Wilkins, Neil Trentham, \& Andrew M. Hopkins]  
{
Stephen M. Wilkins$^{1}$\thanks{E-mail: smw@ast.cam.ac.uk}, Neil Trentham$^{1}$, \& Andrew M. Hopkins$^{2}$ \\
$^1$ Institute of Astronomy, University of Cambridge, Madingley Road, Cambridge, CB3 0HA, United Kingdom\\ 
$^2$ School of Physics, University of Sydney, NSW 2006, Australia
}
\maketitle

\begin{abstract} 
{
We present a compilation of measurements of the stellar mass density as a function of redshift. Using this stellar mass history we obtain a star formation history and compare it to the instantaneous star formation history. For $z<0.7$ there is good agreement between the two star formation histories. At higher redshifts the instantaneous indicators suggest star formation rates larger than that implied by the evolution of the stellar mass density. This discrepancy peaks at $z=3$ where instantaneous indicators suggest a star formation rate around $0.6$ dex higher than those of the best fit to the stellar mass history. We discuss a variety of explanations for this inconsistency, such as inaccurate dust extinction corrections, incorrect measurements of stellar masses and a possible evolution of the stellar initial mass function.
}				   
\end{abstract} 


\section{Introduction}

Much contemporary research in extragalactic astronomy has revolved around the determination of the instantaneous cosmic star formation history (SFH, Madau et al. 1996; Lilly et al. 1996). However, measuring this quantity from observations requires a number of assumptions, with the form of the dust obscuration corrections and stellar initial mass function (IMF, see Kroupa 2007a for a recent overview) being among the most important.

Integration of the instantaneous star formation history over redshift, making appropriate corrections for stellar evolution processes, yields the current stellar mass density. This quantity can be independently measured, typically using extensive galaxy surveys such as the 2dFGRS or SDSS, often combined with near infrared (NIR) measurements. Numerous studies have attempted comparisons of these quantities. Madau, Pozzetti \& Dickinson (1998), Cole et. al (2001), Fontana et al. (2004) and Arnouts et al. (2007) all found good agreement between the SFH with a low dust content and measured values of the stellar mass density. On the other hand there have been a number of studies (Eke et al. 2005; Hopkins \& Beacom 2006, hereafter HB06) which claim that the instantaneous SFH overpredicts the low redshift stellar mass density.
We attempt to investigate this possible discrepancy using a compilation of the most up to date measurements of the stellar mass density history (SMH, the $\Omega_{*}$-redshift relation). This relation is intricately connected to the instantaneous star formation history, but in this context it has some important advantages. The principal advantage is that estimates of stellar mass typically probe a range of the stellar mass function that is somewhat more representative of the stellar mass whereas instantaneous indicators probe only the most massive stars. Furthermore instantaneous measurements can be subject to a greater uncertainty introduced by the effects of dust obscuration.

In \S 2 we present a compilation of both low and high redshift measurements of the stellar mass density. Using these values, in \S 3 we derive a best fitting star formation history. We then compare this estimate in \S 4 to other estimates of the star formation history and highlight any discrepancies. Finally, in \S 5 we present a discussion of our results and end in \S 6 with a summary. Throughout this work, we assume a flat $\Lambda$ CDM cosmology with $\Omega_{\Lambda}=0.7$, $\Omega_{\rm matter}=0.3$ and $H_{0} = 70\,\, {\rm kms}^{-1}\,Mpc^{-1}.$

\section{The Evolution of Stellar Mass}

The stellar mass density at any redshift is obtained by the integral of the galaxy stellar mass function ($\Phi$, GSMF)
\begin{equation}
\rho_{*}=\int_{0}^{\infty}M\Phi(M)dM,
\end{equation} 
with the GSMF defined such that $\Phi(M)dM$ is the density of galaxies with masses between $M$ and $M+dM$.

\begin{landscape}
\setcounter{table}{0}
\begin{table*}
\small
\caption{Summary of recent measurements of the local stellar mass density (top group) and at higher redshifts (bottom group), redshift range, observed mass function range, Salpeter converted value of $\Omega_{*}$ (with original values in parenthesis) and conversion factor to our IMF. $^{a}$Adopts a diet Salpeter IMF. $^{b}$Is independent of an IMF. $^{c}$Adopts a Chabrier IMF. $^{d}$Adopts an IMF with a mass range of $0.1-125 M_{\odot}$. $^{e}$Adopts an alternative cosmology - $h=0.71, \Omega_{m}=0.27, \Omega_{\Lambda}=0.734$. $^{f}$ Adopts an alternative cosmology - $h=0.65, \Omega_{m}=0.35, \Omega_{\Lambda}=0.65$. $^{g}$Published results only include galaxies with the MF range, the authors estimate they obtain $~80\%$ of the mass however, we roughly correct by increasing the estimates by $20\%$. $^{h}$Original mass estimates (in square brackets) include only galaxies with $L_{V}^{rest}>1.4\times10^{10}h^{-2}_{70} L_{\odot}$. The authors state that according to the SDSS luminosity function parameteres they lose $46\%$ of light at $z=0$. At $z=2.8$ due to brightening the authors estimate this becomes $30\%$, thus we correct according to these numbers interpolating inbetween.$^{i}$ Values in square brackets are for the $Z_{\odot}$, one component $\Psi(t)$ model fit with $68\%$ random errors. Utilised values and uncertainties are based on the average of model fits using different metallicities and star formation history models. $^{j}$ Quoted uncertainties are estimates as uncertainties in Fontana et al. (2006) are quoted only for the stellar mass density in the observed range (original values and uncertainties shown in square brackets). $^{k}$ Quoted values are the average of the different methods and samples used in the study and errors are based on the scatter between the these methods and samples. Statistical errors are generally $<10\%$ for each sample and method.}
\begin{tabular}{llcccc}	
 & $z-$range&Observed MF Range &$\Omega_{\rm *;Salpeter}$ ($\Omega_{\rm *;Original IMF}$)& Conversion\\          
 & & ($h_{70}^{-2}\, M_{\odot}$)& & Factor\\          
	  \hline \hline 
{\bf Cole et al. (2001)}               & $z\sim 0$& $10^{9.0}<M/M_{\odot}<10^{11.5}$       & $0.0041 \pm 0.0006$ & 0.58\\

{\bf Bell et al. (2003)}$^{a}$              & $z\sim 0$& $10^{9.0}<M/M_{\odot}<10^{11.5}$       & $0.0040 \pm 0.0012$ ($0.0029 \pm 0.0009$) & 0.58\\

{\bf Panter et al.(2004)}$^{e}$       &$z\sim 0$& $10^{7.5}<M/M_{\odot}<10^{12.0}$       & $0.0034 \pm 0.00011$ & 0.58\\

{\bf Eke et al.  (2005)}              & $z\sim 0$& $10^{8.0}<M/M_{\odot}<10^{12.0}$       & $0.0033 \pm 0.0004$ & 0.58\\

{\bf Sampson et al. (2007)}$^{b}$          & $z\sim 0$& $10^{6.0}<M/M_{\odot}<10^{12.0}$  & $0.0017 \pm 0.0008$& None\\

{\bf Driver et al. (2007)}$^{a}$           & $z\sim 0$& -                                      & $0.0054 \pm 0.0008$ ($0.0039\pm 0.0006 $)& 0.58\\

           \hline 
{\bf Brinchmann \& Ellis (2000)}$^{dfg}$          &$0.20<z<0.50$ & $10^{10.5}<M/M_{\odot}<10^{11.6}$ & $0.0026^{+0.0011}_{-0.0007} $& 0.58\\
                                                  &$0.50<z<0.75$ & $10^{10.5}<M/M_{\odot}<10^{11.6}$ & $0.0029^{+0.0004}_{-0.0008} $& 0.58\\
                                                  &$0.75<z<1.00$ & $10^{10.5}<M/M_{\odot}<10^{11.6}$ & $0.0021^{+0.0008}_{-0.0012} $& 0.59\\

{\bf Rudnick et al. (2003)}$^{h}$                 &$z\sim 0$             & -  & $0.0042^{+0.0005}_{-0.0006} [0.0023^{+0.0002}_{-0.0003}]    $        & 0.58\\ 
SDSS and FIRES                                    &$0.00<z<1.60$  & -  & $0.0017^{+0.0005}_{-0.0005} [0.0010^{+0.0003}_{-0.0003}]    $        & 0.58\\
                                                  &$1.60<z<2.41$  & -  & $0.00037^{+0.00012}_{-0.00016} [0.00022^{+0.00007}_{-0.00010}] $     & 0.59\\
                                                  &$2.41<z<3.20$  & -  & $0.00033^{+0.00010}_{-0.00013} [0.00023^{+0.00007}_{-0.00009}] $     & 0.61\\

{\bf Dickinson et al. (2003)}$^{i}$               &$0.50<z<1.40$  & -  & $0.0023 \pm 0.0006   [0.0021^{+0.0004}_{-0.0004}] $      & 0.58\\
                                                  &$1.40<z<2.00$  & -  & $0.0011 \pm 0.0003   [0.00084^{+0.00040}_{-0.00029}] $   & 0.59\\
                                                  &$2.00<z<2.50$  & -  & $0.00049 \pm 0.00031 [0.00028^{+0.00008}_{-0.00005}] $   & 0.60\\
                                                  &$2.50<z<3.00$  & -  & $0.00038 \pm 0.00022 [0.00024^{+0.00017}_{-0.00009}] $   & 0.61\\

{\bf Fontana et al. (2003)}           &$0.20<z<0.70$  & $10^{10.0}<M/M_{\odot}<10^{11.0}$ & $0.0033  \pm 0.0019 $& 0.58\\
                                                  &$0.70<z<1.30$  & $10^{10.0}<M/M_{\odot}<10^{11.0}$ & $0.0014  \pm 0.0009 $& 0.59\\
                                                  &$1.30<z<2.00$  & $10^{10.0}<M/M_{\odot}<10^{11.0}$ & $0.00058  \pm 0.00034 $& 0.59\\
                                                  &$2.00<z<2.50$  & $10^{10.0}<M/M_{\odot}<10^{11.0}$ & $0.00066  \pm 0.00038 $& 0.60\\
                                                  &$2.50<z<3.20$  & $10^{10.0}<M/M_{\odot}<10^{11.0}$ & $0.00062  \pm 0.00037 $& 0.61\\

{\bf Fontana et al. (2004)}            &$0.20<z<0.70$  & $10^{9.0}<M/M_{\odot}<10^{11.5}$  & $0.0024^{+0.0020}_{-0.0002} $& 0.58 \\ 
                                                  &$0.70<z<1.00$  & $10^{10.0}<M/M_{\odot}<10^{12.0}$ & $0.0023^{+0.0016}_{-0.0002} $& 0.59 \\
                                                  &$1.00<z<1.50$  & $10^{10.5}<M/M_{\odot}<10^{11.5}$ & $0.0012^{+0.0014}_{-0.0002} $& 0.59 \\
                                                  &$1.50<z<2.00$  & $10^{11.0}<M/M_{\odot}<10^{11.5}$ & $0.0011^{+0.0003}_{-0.0008} $& 0.59 \\

\end{tabular}
\end{table*}
\end{landscape}

\begin{landscape}
\begin{table*}
\small
\contcaption{}

\begin{tabular}{llcccc}

 & $z-$range&Observed MF Range &$\Omega_{\rm *;Salpeter}$ ($\Omega_{\rm *;Original IMF}$)& Conversion\\          
 & & ($h_{70}^{-2}\, M_{\odot}$)& & Factor\\          
	  \hline \hline 
{\bf Drory et al. (2004)}                  &$0.40<z<0.60$  & $10^{9.00}<M/M_{\odot}<10^{11.75}$ & $0.0034  \pm 0.0013 $& 0.58 \\
                                                 &$0.60<z<0.80$  & $10^{8.75}<M/M_{\odot}<10^{11.75}$ & $0.0030  \pm 0.0010 $& 0.58 \\
                                                 &$0.80<z<1.00$  & $10^{9.00}<M/M_{\odot}<10^{11.75}$ & $0.0020  \pm 0.0007 $& 0.59 \\
                                                 &$1.00<z<1.20$  & $10^{9.25}<M/M_{\odot}<10^{11.75}$ & $0.0018  \pm 0.0007 $& 0.59 \\

{\bf Drory et al. (2005)}                   &$0.25<z<0.75$  & $10^{8.0}<M/M_{\odot}<10^{11.5}$ & $0.0023  \pm 0.0007  $& 0.58\\
FDF Sample                                       &$0.75<z<1.25$  & $10^{8.5}<M/M_{\odot}<10^{11.5}$ & $0.0019  \pm 0.0006  $& 0.59\\
                                                 &$1.25<z<1.75$  & $10^{9.0}<M/M_{\odot}<10^{11.5}$ & $0.0018  \pm 0.0008  $& 0.59\\
                                                 &$1.75<z<2.25$  & $10^{9.5}<M/M_{\odot}<10^{11.5}$ & $0.00081 \pm 0.00047 $& 0.60\\
                                                 &$2.25<z<3.00$  & $10^{9.5}<M/M_{\odot}<10^{11.5}$ & $0.0010  \pm 0.0006  $& 0.60\\
                                                 &$3.00<z<4.00$  & $10^{9.5}<M/M_{\odot}<10^{11.5}$ & $0.00061 \pm 0.00028 $& 0.61\\
                                                 &$4.00<z<5.00$  & $10^{9.5}<M/M_{\odot}<10^{11.5}$ & $0.00019 \pm 0.00011 $& 0.61\\

GOODS Sample                                     &$0.25<z<0.75$  & $10^{8.0}<M/M_{\odot}<10^{11.5}$ & $0.0024   \pm 0.0012  $& 0.58\\
                                                 &$0.75<z<1.25$  & $10^{8.5}<M/M_{\odot}<10^{11.5}$ & $0.0014   \pm 0.0007  $& 0.59\\
                                                 &$1.25<z<1.75$  & $10^{9.0}<M/M_{\odot}<10^{11.5}$ & $0.00077  \pm 0.00036 $& 0.59\\
                                                 &$1.75<z<2.25$  & $10^{9.5}<M/M_{\odot}<10^{11.5}$ & $0.00088  \pm 0.00047 $& 0.60\\
                                                 &$2.25<z<3.00$  & $10^{9.5}<M/M_{\odot}<10^{11.5}$ & $0.00044  \pm 0.00026 $& 0.60\\
                                                 &$3.00<z<4.00$  & $10^{9.5}<M/M_{\odot}<10^{11.5}$ & $0.00034  \pm 0.00020 $& 0.61\\
                                                 &$4.00<z<5.00$  & $10^{9.5}<M/M_{\odot}<10^{11.5}$ & $0.00017  \pm 0.00014 $& 0.61\\

{\bf Fontana et al. (2006)}$^{j}$                &$0.40<z<0.60$  & $10^{9.0}<M/M_{\odot}<10^{11.5}$  & $0.0021\pm 0.0002 [0.0015 \pm 0.0001]$  & 0.58 \\
                                                 &$0.60<z<0.80$  & $10^{9.0}<M/M_{\odot}<10^{11.5}$  & $0.0017\pm 0.0002 [0.0025 \pm 0.0001]$  & 0.58 \\
                                                 &$0.80<z<1.00$  & $10^{9.5}<M/M_{\odot}<10^{11.5}$  & $0.0014\pm 0.0002 [0.0011 \pm 0.0001]$  & 0.59 \\
                                                 &$1.00<z<1.30$  & $10^{9.5}<M/M_{\odot}<10^{11.5}$  & $0.0011\pm 0.0002 [0.0013 \pm 0.0001] $  & 0.60 \\
                                                 &$1.30<z<1.60$  & $10^{10.0}<M/M_{\odot}<10^{11.5}$ & $0.00086\pm 0.0008 [0.00067 \pm 0.00005]$ & 0.60 \\
                                                 &$1.60<z<2.00$  & $10^{10.0}<M/M_{\odot}<10^{11.5}$ & $0.00064\pm 0.0009 [0.00058 \pm 0.00006]$ & 0.60 \\
                                                 &$2.00<z<3.00$  & $10^{10.3}<M/M_{\odot}<10^{11.5}$ & $0.00035\pm 0.0005 [0.00029 \pm 0.00003]$ & 0.60 \\
                                                 &$3.00<z<4.00$  & $10^{10.5}<M/M_{\odot}<10^{11.5}$ & $0.00014\pm 0.0006 [0.00012 \pm 0.00004 ]$ & 0.61 \\

{\bf Arnouts et al. (2007)}                      &$0.2<z<0.4$    & -                                 & $0.0046  \pm 0.0019  $& 0.58\\
                                                 &$0.4<z<0.6$    & -                                 & $0.0035  \pm 0.0014  $& 0.58\\
                                                 &$0.6<z<0.8$    & -                                 & $0.0033  \pm 0.0014  $& 0.59\\
                                                 &$0.8<z<1.0$    & -                                 & $0.0039  \pm 0.0016  $& 0.59\\
                                                 &$1.0<z<1.2$    & -                                 & $0.0027  \pm 0.0011  $& 0.59\\
                                                 &$1.2<z<1.5$    & -                                 & $0.0019  \pm 0.0007  $& 0.59\\
                                                 &$1.5<z<2.0$    & -                                 & $0.0011  \pm 0.0004  $& 0.60\\
{\bf Pozzetti et al. (2007)}$^{ck}$               &$0.05<z<0.40$  & $10^{7.0}<M/M_{\odot}<10^{11.5}$ & $0.0036 \pm 0.0008$ ($0.0021 \pm 0.0005$) & 0.58 \\
                                                 &$0.40<z<0.70$  & $10^{8.5}<M/M_{\odot}<10^{11.5}$  & $0.0027 \pm 0.0005$ ($0.0016 \pm 0.0003$) & 0.58 \\
                                                 &$0.70<z<0.90$  & $10^{9.0}<M/M_{\odot}<10^{11.5}$  & $0.0020 \pm 0.0007$ ($0.0012 \pm 0.0004$) & 0.59 \\
                                                 &$0.90<z<1.20$  & $10^{9.5}<M/M_{\odot}<10^{11.5}$  & $0.0017 \pm 0.0005$ ($0.0010 \pm 0.0003$) & 0.59 \\
                                                 &$1.20<z<1.60$  & $10^{9.5}<M/M_{\odot}<10^{11.5}$  & $0.0014 \pm 0.0005$ ($0.0008 \pm 0.0003$) & 0.59 \\
                                                 &$1.60<z<2.50$  & $10^{10.0}<M/M_{\odot}<10^{11.5}$ & $0.0014 \pm 0.0004$ ($0.0008 \pm 0.0002$) & 0.60 \\

{\bf Bell et al. (2007)}$^{c}$                   &$0.20<z<0.40$  & $10^{9.0}<M/M_{\odot}<10^{11.5}$  & $0.0026 \pm 0.0007$ ($0.0015 \pm 0.0004$) & 0.58 \\             
                                                 &$0.40<z<0.60$  & $10^{9.0}<M/M_{\odot}<10^{11.5}$  & $0.0025 \pm 0.0002$ ($0.0014 \pm 0.0001$) & 0.58 \\
                                                 &$0.60<z<0.80$  & $10^{9.0}<M/M_{\odot}<10^{12.0}$  & $0.0026 \pm 0.0003$ ($0.0015 \pm 0.0002$) & 0.59 \\
                                                 &$0.80<z<1.00$  & $10^{9.0}<M/M_{\odot}<10^{12.0}$  & $0.0018 \pm 0.0005$ ($0.0011 \pm 0.0003$) & 0.59 \\

\end{tabular}
\end{table*} 
\end{landscape}

\subsection{Stellar Mass Density Measurements at $z=0$}

The most prevalent method of estimating stellar masses is to obtain the closest fit between observed galaxy spectral energy distributions (SEDs) and a library of template SEDs. Template SEDs are typically have a range of star formation histories, metallicity distributions and dust content. These are typically created using a population synthesis model, a code which generates the SED of a stellar population by combining a stellar evolution prescription with a library of stellar spectra. In addition to requiring a star formation history, metallicity distribution and dust content population synthesis models also require the assumption of a specific form of the initial mass function (IMF). Variations between alternative IMFs can for a given star formation history produce different SEDs, and thus different recovered masses making this technique then IMF dependent. Comparing results from different studies can then only be accurately achieved by converting to the same IMF. Furthermore, different population synthesis models often use different evolution prescriptions and spectral libraries. This in turn can produce variations in the SEDs (Bruzual \& Charlot 2003) of template galaxies and thus a contrasting recovered mass (Panter et al. 2007; Pozzetti et al. 2007).

An early major study using this technique is that of Cole et al. (2001) who use galaxy spectra from the 2dFGRS and the NIR luminosities from two Micron All Sky Survey (2MASS). They find $\Omega_{*}(z=0)=0.0041\pm 0.0006$, where $\Omega_{*}$ is the stellar mass density in units of the critical density assuming the Salpeter IMF. Eke et al. (2005) use an updated version of these catalogues to obtain $\Omega_{*}=0.0033\pm 0.00014$ similarly assuming a Salpeter IMF but using a more generalised set of template star formation histories. Bell et al. (2003) also use the 2MASS catalogue but instead of the 2dFGRS use the early data release of the SDSS. They obtain a stellar mass density of $\Omega_{*}=0.0028\pm 0.00086 $, using a modified Salpeter IMF (diet Salpeter, Bell \& de Jong 2001), a Salpeter IMF with a lower limit of $M=0.188 M_{\odot}$ motivated by numerous observations which suggest the Salpeter IMF overpredicts the number of low mass stars (Kroupa 2007a). Panter, Heavens \& Jimenez (2004) use the SDSS data release 1 obtaining a result of $\Omega_{*}=0.0034\pm 0.00011$ for the Salpeter IMF. Driver et al. (2007) adopt the diet Salpeter IMF and using a survey of galaxies in the $B$-band obtain a result of $\Omega_{*}=0.0054 \pm 0.0008$. 

Sampson et al. (2007) determine the local stellar mass independent of the IMF, updating the work of Read \& Trentham (2005).  This is achieved by constructing the local type-specific $R$-band luminosity function from a variety of local galaxy catalogues. Type and luminosity specific mass-to-light ratios obtained from dynamical and lensing measurements, which do not depend strongly on the IMF are then used to determine the GSMF. Integration over this GSMF yields $\Omega_{*}=0.0017\pm 0.0008$. Although this technique is independent of the IMF it does suffer a variety of other limiting assumptions, principally the characteristics of the dark matter component. 
In this paper we choose to adopt an IMF consistent with recent observations of the local IMF (Kroupa 2007a). We define the IMF a power law with an index of $\alpha=1.0$ over the range $0.1<M/M_{\odot}<0.5$ and the Salpeter index ($\alpha=2.35$) over the range $0.5<M/M_{\odot}<100$. Converting estimates using alternative IMFs to this IMF is achieved using a population synthesis code such as the code Bruzual \& Charlot (2003) or PEGASE (Fioc \& Rocca-Volmerange 1997). Because of the ease of introducing alternative IMFs that it allows we adopt the PEGASE.2 code. We find that the choice of this IMF implies significantly smaller stellar masses than the Salpeter IMF, that is Mass(Our IMF) $\sim$ Mass(Salpeter) $-0.23\,$dex at $z=0$. This conversion factor mildly evolves increasing to {\em Mass(Our IMF)=Mass(Salpeter)-0.21$\,$} dex at $z=4$.
 
\subsection{Stellar Mass Density at Higher Redshifts}
                              
\begin{figure*}
\includegraphics[width=40pc]{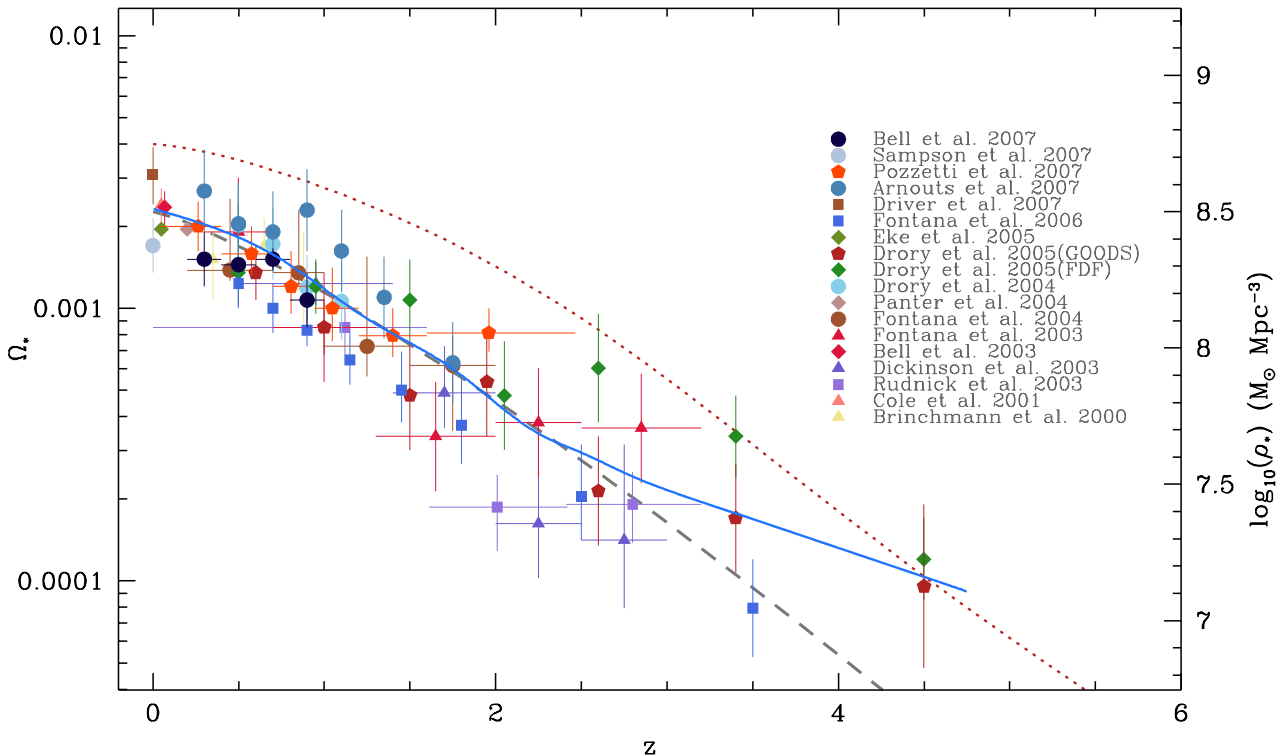} 
\caption{Stellar mass as a function of redshift converted to our IMF and cosmology. The solid blue line is created by a simple weighted bining procedure and the dashed grey line is a best fit with the parameterisation  $\rho_{*}(z)=a\,\times\,e^{-bz^{c}}$ and $a=0.0023$, $b=0.68$ and $c=1.2$. The dotted red line is the prediction of HB06 generated by integrating the observed instantaneous star formation history.}
\end{figure*}

Similar techniques as those utilised at $z\sim 0$ have recently been applied to surveys of the higher redshift galaxy population. The main difference is that surveys at higher redshift often have far fewer galaxies than those locally (such as SDSS and 2dFGRS) and thus may be incomplete at the extremes stellar mass function. To account for this the mass function is often extrapolated beyond the observed range to cover a range of typically $10^{8}<M/M_{\odot}<10^{13}$. This is most often achieved by fitting a Schechter (1976) function.

A compilation of both local and high redshift studies is presented in table 1. In this table measurements are converted to the Salpeter IMF with the additional conversion to our IMF explicitly stated. The measurements converted to our IMF are shown in figure 1. The solid line in figure 1 was generated by binning each measurement weighted by its uncertainty. A parametric form, $\rho_{*}(z)=a\,\times\,e^{-bz^{c}}$ is also shown (dashed line) with best fitting parameters $a=0.0023$, $b=0.68$ and $c=1.2$. Although this parametric fit works well up to intermediate redshifts, at $z\sim 4$ it is significantly smaller than the data.

\section{Reconstructing the star formation history}
The stellar mass density history $\rho_{*}(t)$ can be expressed as the integral of the star formation history $\dot{\rho_{*}}(t)$ corrected for the effects of mass loss through stellar evolution processes such as supernovae and stellar winds (Woosley \& Weaver 1995,  Renzini \& Voli 1981). Often this is expressed as $\rho_{*}(t)=(1-R)\int_{0}^{t}\dot{\rho}_{*}dt'$, where $R$ represents the fraction of material returned to the ISM. In reality this is an approximation as $R$ is a product of the previous star formation history. For example a recent burst of star formation would have returned only small amount material where as a burst at higher redshift would have returned significantly more material by the current epoch. To account for this, the stellar mass history can be expressed more accurately as
\begin{equation}
\rho_{*}(t)=\int_{0}^{t}\dot{\rho}_{*}(t')(1-f_{{\rm r}}[t-t']) dt',
\end{equation}
where $f_{{\rm r}}[t-t']$ is the fraction of stellar mass created at $t'$ that has been returned to the ISM by $t$. This quantity can be calculated by considering the mass evolution of an instantaneous burst of star formation using either population synthesis codes or analytical formulae derived to fit the initial-mass/final-mass relations from observations such as those of Woosley \& Weaver (1995) and Renzini \& Voli (1981). We obtain this quantity using the PEGASE code to model the mass loss characteristics after a burst of star formation. We check the results of this by comparing to both the analytic formulae of Hurley, Pols \& Tout (2000) as well the alternative population synthesis code of Bruzual \& Charlot (2003). In the former case we find good agreement for $f_{r}$ for our IMF. Similarly we find that both PEGASE and Bruzual \& Charlot (2003) provide similar results for both the Salpeter and Chabrier IMF.

Inverting equation 2 can be used to determine the star formation history from the observed evolution of stellar mass. This yields
\begin{equation}
\dot{\rho}_{*}(t)=d\rho_{*}(t)/dt + d\rho_{*;r}(t)/dt,
\end{equation}
where $d{\rho}_{*}(t)/dt$ is the time derivative of the observed stellar mass history (i.e. that presented in figure 1) and $\dot{\rho}_{*;r}(t)$ is the rate at which material is returned to the ISM. This is a function of the previous star formation history and the returned function introduced above, 
\begin{equation}
\rho_{*;r}(t)=\int_{0}^{t}\dot{\rho}_{*}(t')(f_{r}[t-t'])dt'.
\end{equation}

Using the compilation presented in \S 2 and the formalism presented in equations 3 and 4 a best fit star formation history along with $1\sigma$ and $3\sigma$ uncertainty regions was generated. For future ease of comparison we follow the lead of other authors and express our star formation history by the parameterisation of Cole et al. (2001), $\dot{\rho}_{*}=(a+bz)h/[1+(z/c)^{d}]$. We find best fitting parameters of $a=0.014$, $b=0.11$, $c=1.4$, and $d=2.2$ for $h=0.7$. This best fit star formation history, and the associated $1\sigma$ and $3\sigma$ uncertainty regions are shown in figure 2.

\section{Comparison with other indicators of Star Formation}

\begin{figure*}
\includegraphics[width=35pc]{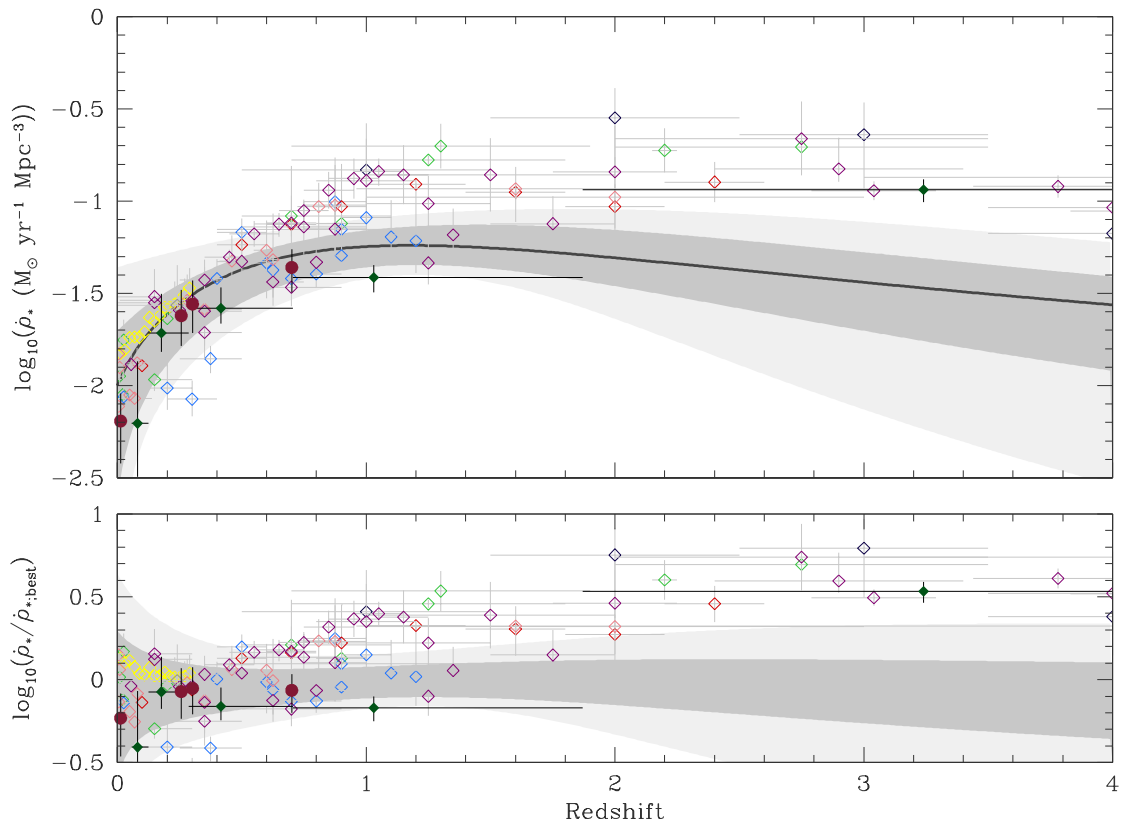} 
\caption{Comparison of the star formation history inferred from the stellar mass density history compared to other measurements of star formation rates. The star formation history inferred from the evolution of stellar mass is shown by the $1\sigma$ and $3\sigma$ uncertainty regions (dark and light grey shaded areas respectively). The dark solid line is the parameterised best fit to our star formation history discussed in the text. The lower panel displays the ratio of this best fit SFH to the other measurements. Fossil History results are from Panter (2006) (dark green). Instantaneous indicators of star formation are open diamonds colour coded by type and are dust corrected as laid out by Hopkins \& Beacom (2006). [OII] lines (blue): Teplitz et al. (2003); Gallego et al. (2002); Hogg et al. (1998); Hammer et al. (1997).  Radio/Sub-mm/FIR (red): Flores et al. (1999); Barger, Cowie \& Richards (2000); Condon, Cotton, \& Broderick (2002); Sadler et al. (2002); Serjeant, Gruppioni, \& Oliver (2002); Machalski \& Godlowski (2000); Haarsma et al. (2000); Condon (1989); P{\'e}rez-Gonz{\'a}lez et al. (2005). UV indicators (purple): Giavalisco et al. (2004); Wilson et al. (2002); Massarotti, Iovino, \& Buzzoni (2001); Sullivan et al. (2000); Steidel et al. (1999);  Cowie, Songaila , \& Barger (1999); Treyer et al. (1998); Connolly et al. (1997); Lilly et al. (1996); Schiminovich et al (2005); Wolf et al (2003). Xray (grey) Georgakakis et al. (2003). SDSS data (yellow) Baldry et al (2005). H$\alpha$ and H$\beta$ (green): Pettini et al. (1998); P{\'e}rez-Gonz{\'a}lez et al. (2003); Tresse et al. (2002); Moorwood et al. (2000); Hopkins, Connolly, \& Szalay (2000); Sullivan et al. (2000); Glazebrook et al. (1999); Yan, Windhorst, \& Cohen (1999); Tresse \& Maddox (1998); Gallego et al. (1995); Hanish et al (2006). Supernovae Rates are filled red circles from Dahlen et al. (2004) and Cappellaro et al. (2005)}
\end{figure*}

\subsection{Instantaneous Indicators of Star Formation}

Star formation rates are most often calculated from instantaneous indicators (see Kennicutt 1998 or Calzetti 2007 for an overview). This is typically emission associated with ongoing star formation. Since very massive stars have lifetimes which are short compared to typical star formation event timescales, emission associated with these stars is used as an indicator of their formation rate. These young, massive stars completely dominate the integrated UV emission of a galaxy, which is thus used as an indicator for their presence. Emission from these stars, particularly that which is shortward of the Lyman limit is reprocessed by hydrogen to produce nebular lines such as H$\alpha$ and H$\beta$. These along with other indicators such as forbidden lines and infrared emission (in the case of heavily obscured star formation regions) can be used as tracers of the formation of massive stars. Extrapolating down the IMF then yields the total star formation rate. Using the compilation of HB06 (an updated version of Hopkins 2004) we compare a range of instantaneous measurements to our star formation history, shown in figure 2. The measurements are corrected for the effect of dust attenuation using a common obscuration correction (as in HB06). As detailed in Hopkins (2004), the common obscuration correction corrects emission-line measurements using $A_{\rm H\alpha}= 1.0$ and the Cardelli et al. (1989) Galactic obscuration curve, and corrects continuum UV measurements assuming $A_{V_{\rm star}}= 0.52$ and the Calzetti et al. (2000) starburst obscuration curve. We convert from a Salpeter IMF to our adopted IMF by scaling the values by $-0.15\,$ dex obtained from the ratio of UV luminosities for a simple burst using the PEGASE code. Below $z=0.7$ the instantaneous star formation rate measurements show very good agreement with our star formation history. At higher redshifts however, we see an increasing systematic deviation, with instantaneous indicators suggesting much larger star formation rates. This deviation appears to peak $z~3$ where the best fit to instantaneous measurements is approximately $4$ times larger than the best fit to our star formation history. 

\subsection{Core Collapse Supernovae Rates}
A similar estimate of the star formation history can be obtained from the density rate of core collapse supernovae (CCSN; Including Type II, Ib and Ic supernova). These are produced by very massive stars $8-50\, M_{\odot}$, at the end of their lifetime (Fryer 1999,Dahlen et al. 2004). Because these stars are shortlived compared to typical lengths of star formation episodes they are contemporary and can thus be used as a probe of the instantaneous SFH in a similar way to the photometric estimates described in \S 4.1. Assuming that all the stars in this range undergo a CCSN, the rate is related to the star formation rate through (Dahlen et al. 2004; Fryer 1999; Madau, della Valle, \& Panagia 1998), 
\begin{equation}
\dot{\rho}_{*}(z)=\frac{\int_{0.1}^{101}M\Psi(M){\rm d}M}{\int_{8}^{50}\Psi(M){\rm d}M}\,\dot{\rho}_{CCSN}(z).
\end{equation}
For our choice of IMF this becomes $\dot{\rho}_{*}(z)=100\,(M_{\odot})\,\dot{\rho}_{CCSN}(z)$. Observations of CCSN rates converted into star formation rate densities in this way are shown in figure 2. These measurements are completely consistent with both the $1\sigma$ uncertainty region of our star formation history and that of instantaneous indicators. CCSN rate measurements are however presently limited to less than $z<1$.

\subsection{Star Formation Rates from the Fossil Record}
An alternative estimate of the star formation history can be generated from the fossil record of star formation in nearby galaxies. This involves determining the distribution of stellar ages for individual galaxies through the use of stellar population synthesis models. Heavens et al. 2004 carried out such an analysis on the SDSS data release 1. A similar analysis was carried out by Panter et al. (2007) using an updated version of the SDSS catalogue (SDSS data release 3) and higher resolution model spectra. The results of this updated version of the analysis are shown in figure 2.

\section{Results}

For $z<0.7$ the $1\sigma$ uncertainty region of our SFH is consistent with best fit instantaneous star formation history obtained by HB06. This is consistent with the findings of recent study of Bell et al. (2007) (a study included in our compilation) who find that the instantaneous star formation history (from Hopkins 2004) correctly matches the assembly of stellar mass over this redshift range.
 
For $z>0.7$ the best fit star formation history of HB06 is consistently higher than our star formation history. At $z=3$ the best fit of HB06 SFH implies a star formation rate around $4$ times ($0.6\,$ dex) larger than that inferred from the stellar mass density. This large deviation at high redshift offers an explanation for why the integrated star formation history implies a local stellar mass density in excess of that measured.

This leads to two possible conclusions, both of which may be correct to some degree. Either stellar mass estimates are incorrect or the star formation history of HB06 is overestimated at high redshifts ($z>0.7$). The former possibility requires that both high and low redshift measurements of the stellar mass density are incorrect suggesting a systematic underestimation. Such an underestimation could be caused by a number of possibilities, principally including incorrect extrapolation outside the measured GSMF or erroneous calculation of stellar masses. The latter possibility could be due to the often limited extent of photometric information (however Fontanna et al. 2006 investigated the effect of including Spitzer infrared information on stellar mass determination for high redshift galaxies and found little change) or some other systematic effect such as metallicity-age degeneracies or the effect of dust. Other possibilities include the presence of stellar mass missed from surveys in extended galactic halos (discussed by Bernstein, Freedman, \& Madore 2002ab) or as free stars in clusters.

If the instantaneous SFH has been overestimated at high redshifts, this could also resolve the issue. A common explanation leading to this conclusion is that the effect of dust attenuation has been overestimated in measurements such as those presented in HB06. Arnouts et al. (2007) for example found that using a milder dust attenuation correction stellar mass measurements and predictions from instantaneous star formation rates can be reconciled up to at least $z=1.75$. The stellar mass estimates of Arnouts et al. (2007) are, however, significantly above the average of the compilation, as can be seen in figure 1. This suggests that even this mild form of dust correction will still not be consistent with the average of the compilation. The dust correction applied over $3\lesssim z \lesssim 6$ by HB06 varies between $2.4$ to $3.7$ (depending on rest-frame UV wavelength). Reconciling the HB06 SFH with that inferred from the stellar mass density would require scaling down by almost the same factors over this redshift range, implying that there be little or no dust obscuration at all over this redshift range. This is inconsistent with the recent results of Ouchi et al. (2004), Ando et al. (2005), Chary et al. (2005) and Reddy et al. (2007), which show that at least some samples of high redshift UV/optically selected star forming galaxies contain significant quantities of dust. As noted by many authors (e.g., Afonso et al. 2003), UV/optical selection is strongly biased against dusty systems, so the presence of such obscuration even in UV-selected samples suggests that the extent of the obscuration corrections utilised by HB06 is unlikely to be overestimated. It is certainly not overestimated by an amount sufficient to fully resolve the discrepancy between the observed SFH and that inferred from the stellar mass density.

An alternative solution is that the larger star formation rates could be explained by an evolution of the star formation rate calibration (i.e. the conversion factor for converting a UV luminosity density in a star formation rate density). Such a process could speculatively be driven by an evolution of the IMF. Investigation of an environmental or temporal evolution of the IMF has been carried out by a number of authors. Resolved studies have found little variation of the IMF (Kroupa 2007b) with either metallicity or environment, however these have generally been limited to local galaxies and as such may not be fully representative of the entire galaxy population. The possibility of alternative IMFs in galaxies with apparently very large star formation rates is also discussed in Nagashima et al. (2005), Baugh et al. (2005), Le Delliou et al. (2006), Lacey et al. (2007) who suggest a flat IMF to reconcile a number of observations with the predictions of the GALFORM (Cole et al. 2000) semi-analytical model. Fardal et al. (2007) also investigates the possibility of a different IMF in starburst galaxies to explain a discrepancy between the extragalactic background light, the instantaneous star formation history and the $K$-band luminosity density.

Within the context of the deviation we observe, an IMF which produces more emission associated with instantaneous indicators (such as the UV or H$\alpha$ luminosity) per unit mass created is required at high redshift. Such an IMF is likely to be top-heavy (or high mass biased) since instantaneous indicators are typically dominated by very massive stars.

In order to investigate this we introduce a simple composite IMF which has an implicit time dependence, $\xi(m,t)=\xi_{1}(m)+b(t)\,\xi_{\rm HMB}(m)$. Here $\xi_{1}(m)$ is our normal IMF and $\xi_{\rm HMB}(m)$ is a simple high-mass-biased IMF consisting of Salpeter slope truncated at $5 M_{\odot}$. Introducing this evolving IMF changes both the star formation history implied by instantaneous indicators, the fraction of material recycled as a function of age and the observed stellar mass density. 

Using a simple model for the evolution of the IMF of $b(t)=1-e^{-\tau z}$ with $\tau=0.6$ we find significantly increased agreement, shown in figure 3. In the top panel the best fit to the HB06 star formation history for a non-evolving (dotted line) and evolving (solid line) IMF is shown alongside the star formation history implied by the evolving IMF corrected evolution of stellar mass. In the bottom panel the evolution of the stellar mass density predicted from the best fit star formation history of HB06 for non-evolving and evolving IMFs is also shown. The open and filled points denote the measurements of the stellar mass density for a non-evolving and evolving IMF respectively. In both panels it is clear that the inclusion of this evolving IMF reduces the discrepancy. 

This is a simple and ad hoc model, and many other forms of evolving IMF may also reproduce the relationship between the instantaneous SFH and the SFH derived from the evolution of stellar mass. We use this model here simply to illustrate the effect of the increasing high mass bias of an IMF toward high redshift, and to show that such an effect is sufficient to explain the discrepancies between the instantaneous SFH and the SFH inferred from the SMH. In reality it is likely that $\xi(m,t)$ has a complex evolution and this is currently being investigated in some detail in ongoing work.

\begin{figure}
\includegraphics[width=20pc]{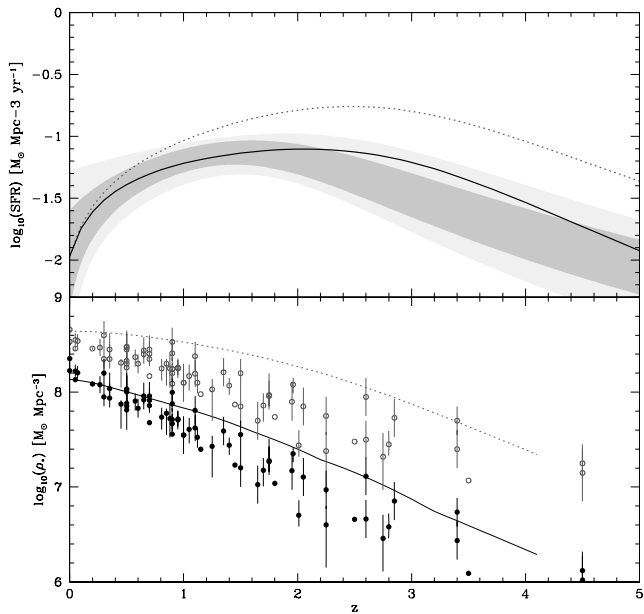} 
\caption {Top panel - The best fit instantaneous star formation history to the compilation of HB06 obtained with a universal IMF (dotted line) compared with that obtained with an evolving IMF (solid line). The shaded regions represent the $1\sigma$ and $3\sigma$ uncertainty regions of the star formation history inferred from the evolvution of the average stellar mass density assuming an evolving IMF. Bottom panel - the observed stellar mass density assuming a universal IMF (open circles) and assuming an evolving IMF (filled circles) as well as the predictions from the instantaneous star formation history assuming a universal IMF (dotted line) and evolving IMF (solid line).}
\label{fig:zd1}
\end{figure}

\section{Summary \& Discussion}

In this work a compilation of stellar mass density measurements over the range $0<z<4$ was compiled and converted into a modified form of the Salpeter IMF. A simple analytical parameterisation of the form $\rho_{*}(z)=a\,\times\,e^{-bz^{c}}$ with $a=0.0023$, $b=0.68$ and $c=1.2$ was found to fit the observations well. Using this compilation a best fitting star formation history was the obtained. This star formation history is well described by the Cole et al. (2001) parameterisation with $a=0.014$, $b=0.11$, $c=1.4$, and $d=2.2$. This star formation history was compared to other indicators of the star formation history including instantaneous measures, the observed rates of core collapse supernovae and the results of recent analysis of the fossil record from the SDSS. Below a redshift of $0.7$  there is good agreement between these estimates and the $1\sigma$ uncertainty region of our star formation history. At progressively higher redshifts, the stellar mass density and instantaneous indicator inferred star formation histories become inconsistent. Instantaneous measures at $z=3$ imply best fit star formation rates $4$ times larger than those inferred from the stellar mass density. 

There are a number of possible causes of this tension. These include principally, uncertainty in the effects of dust on both stellar mass estimates and high redshift star formation rate estimates as well as modeling uncertainties and issues regarding the completeness of the galaxy stellar mass function.

In addition, there are more speculative solutions such as an effective temporal evolution of the initial mass function. We have identified a simple, non-unique model for an evolving IMF that reconciles both the SFH and stellar mass history. Other recent evidence for an evolving IMF has been explored by Dav{\'e} (2008), and Van Dokkum (2008), who provide different parameterisations. A more vigorous investigation of the implications of an evolving IMF, and suitable choices of parameterisation, is currently underway.

Given the importance of both stellar mass density and star formation rate density measurements to our understanding of the galaxy formation process it is crucial that this discrepancy be resolved. To achieve this, improvements need to be made to measurements, and extensions, such as an evolving IMF, to galaxy formation models need to be implemented. Specifically measurements of the stellar mass density can be improved through the use of larger and deeper surveys to minimise incompleteness, a deeper understanding of the effect of dust attenuation and improvements in both the template fitting procedure and the underlying population synthesis models used to create the templates. Improvements in the star formation history can also be made by refining models of dust attenuation and using larger deeper surveys. Further constraints on the star formation history can also come from improvements in core collapse supernovae rates studies, and more detailed observations of the extragalactic background light, diffuse supernovae neutrino background and the chemical evolution.

\vskip 20pt

\noindent We would like to thank the referee Lucia Pozzetti for many helpful comments during the preparation of this document. We would also like to thank again Lucia Pozzetti as well as Leda Sampson and Niv Drory for the provision of an electronic version of their stellar mass density estimates and Ben Panter for providing us with star formation rate estimates. SMW acknowledges support of an STFC studentship and of King's College, NT acknowledges support provided by STFC and AMH acknowledges support provided by the Australian Research Council in the form of a QEII Fellowship (DP0557850).

\end{document}